\documentclass[11pt]{article}
\usepackage[margin=1in]{geometry}
\usepackage[T1]{fontenc}
\usepackage[utf8]{inputenc}
\usepackage{lmodern}
\usepackage{amsmath,amsthm,amssymb}
\usepackage{newunicodechar}
\newunicodechar{⟨}{\ensuremath{\langle}}
\newunicodechar{⟩}{\ensuremath{\rangle}}
\newunicodechar{⊑}{\ensuremath{\sqsubseteq}}
\newunicodechar{⊓}{\ensuremath{\sqcap}}
\newunicodechar{⊔}{\ensuremath{\sqcup}}
\newunicodechar{⊥}{\ensuremath{\bot}}
\newunicodechar{⊤}{\ensuremath{\top}}
\newunicodechar{Φ}{\ensuremath{\Phi}}
\newunicodechar{Ψ}{\ensuremath{\Psi}}
\newunicodechar{Ω}{\ensuremath{\Omega}}
\newunicodechar{Δ}{\ensuremath{\Delta}}
\newunicodechar{τ}{\ensuremath{\tau}}
\newunicodechar{α}{\ensuremath{\alpha}}
\newunicodechar{γ}{\ensuremath{\gamma}}
\newunicodechar{μ}{\ensuremath{\mu}}
\newunicodechar{σ}{\ensuremath{\sigma}}
\newunicodechar{§}{\S}
\newunicodechar{±}{\ensuremath{\pm}}
\newunicodechar{·}{\ensuremath{\cdot}}
\newunicodechar{¹}{\textsuperscript{1}}
\newunicodechar{⁰}{\textsuperscript{0}}
\newunicodechar{⁵}{\textsuperscript{5}}
\newunicodechar{⁶}{\textsuperscript{6}}
\newunicodechar{⁹}{\textsuperscript{9}}
\newunicodechar{⁻}{\textsuperscript{-}}
\newunicodechar{₁}{\textsubscript{1}}
\newunicodechar{₂}{\textsubscript{2}}
\newunicodechar{–}{--}
\newunicodechar{—}{---}
\newunicodechar{′}{\ensuremath{\prime}}
\newunicodechar{→}{\ensuremath{\to}}
\newunicodechar{∅}{\ensuremath{\emptyset}}
\newunicodechar{∈}{\ensuremath{\in}}
\newunicodechar{−}{\ensuremath{-}}
\newunicodechar{∘}{\ensuremath{\circ}}
\newunicodechar{√}{\ensuremath{\sqrt{\vphantom{x}}}}
\newunicodechar{∩}{\ensuremath{\cap}}
\newunicodechar{≈}{\ensuremath{\approx}}
\newunicodechar{≠}{\ensuremath{\neq}}
\newunicodechar{≥}{\ensuremath{\geq}}
\newunicodechar{×}{\ensuremath{\times}}
\newunicodechar{□}{\ensuremath{\square}}
\usepackage{array,booktabs,graphicx,hyperref,microtype,xcolor,colortbl,enumitem,titlesec,fancyhdr,caption,tikz,pgfplots,tabularx,float,fvextra}
\usetikzlibrary{arrows.meta,positioning,calc,fit,backgrounds}
\usepgfplotslibrary{groupplots}
\pgfplotsset{compat=1.18}
\newcolumntype{Y}{>{\raggedright\arraybackslash}X}
\setlist[itemize]{leftmargin=1.5em,topsep=0.3em,itemsep=0.2em,parsep=0pt}
\setlist[enumerate]{leftmargin=1.7em,topsep=0.3em,itemsep=0.2em,parsep=0pt}
\titleformat{\section}{\large\bfseries}{\thesection}{0.5em}{}
\titleformat{\subsection}{\normalsize\bfseries}{\thesubsection}{0.5em}{}
\titleformat{\subsubsection}{\normalsize\itshape}{\thesubsubsection}{0.5em}{}
\titlespacing*{\section}{0pt}{1.2ex plus 0.4ex minus 0.2ex}{0.6ex}
\titlespacing*{\subsection}{0pt}{1.0ex plus 0.3ex minus 0.2ex}{0.4ex}
\titlespacing*{\subsubsection}{0pt}{0.8ex plus 0.2ex minus 0.1ex}{0.3ex}
\title{Ternary Memristive Logic: Hardware for Reasoning Realized via Domain Algebra}
\author{\textbf{Chao Li}\\Deepleap.ai (lichao@deepleap.ai)}
\date{}
\begin{document}
\frenchspacing
\maketitle
\begin{abstract}
Memristive crossbar arrays perform matrix-vector multiplication in memory, eliminating data movement between processor and storage. But each junction stores a numerical weight — a fragment that only acquires meaning when aggregated with an entire row, passed through activation circuits, and decoded by external logic. The junction alone answers nothing.

We show a fundamentally different use of the same hardware. The CDC framework (Li, Wang \& Zhao, 2026a,b; Li \& Wang, 2026c) establishes that when domain is embedded in predicate arity, the resulting four-tuple ⟨c, r, c′, d⟩ forms a computable Heyting algebra with typed Galois connections, and reasoning collapses into CRUD operations on stored data. This paper provides the physical counterpart: each memristive junction stores a \textbf{complete, self-contained CDC assertion} — "this medical relationship holds (or is negated, or is undefined) in this specific clinical domain." Reading one junction answers one clinical question. No aggregation, no activation function, no decoder.

The core contribution is not the ternary state mapping — three resistance levels are well known. The contribution is that \textbf{a compile-time snapshot of the CDC domain algebra maps structure-preservingly onto a memristive crossbar topology}. Given the algebraic results of Li, Wang \& Zhao (2026b) — Heyting algebra structure, typed Galois connection, monotone closure — we define a mapping Φ from a materialized domain algebra DA* to a crossbar topology CT and show Φ preserves each of these structures. Fiber isolation maps to array isolation. The specialization order maps to directed inter-array wiring. The typing function τ maps (after unfolding across the inheritance graph) to meta-controlled transistor gates. Cross-fiber bridges map to cross-reference registers. The closure property maps to physical idempotence of the inheritance cycle.

This homomorphism is what distinguishes the design from a ternary content-addressable memory (TCAM). A TCAM stores values in a flat array; entries can be rearranged without changing behavior. In the CDC crossbar topology, the physical wiring IS the domain algebra — rearranging connections changes the reasoning semantics. \textbf{A TCAM stores values. This chip stores an algebra.}

We present a concrete design for an ICD-11 respiratory disease classification chip (1,247 entities, three classification axes, $\approx$ 136,000 junctions) realized with 1T1R cells to eliminate sneak paths; specify the cascade mechanism that performs transitive classification through a row-column feedback path; introduce cross-reference registers for bridge operations between parallel classification axes; and validate the five core reasoning capabilities plus cross-axis bridge operations with a behavioral simulation whose results, at nominal device parameters ($\sigma_{\log}=0.15$, SNR = 20 dB), are error-free across 100,000 trials per test — with a substantial safety margin before the design degrades. Li and Wang (2026c) demonstrated that representation-computation unity dissolves the storage-computation boundary in software. This paper demonstrates the same dissolution in physics.

\end{abstract}
\noindent\textbf{Keywords:} Memristive Computing, Structural Homomorphism, Domain Algebra, Ternary Logic, ICD-11, Medical AI, Knowledge Representation, 1T1R

\section{Two Kinds of "In-Memory Computing"}

\subsection{The Neural Network Kind: Junctions as Fragments}

A memristive neural network accelerator stores a weight matrix $W$ in a crossbar array. Each junction $(i, j)$ holds a conductance value $G_{ij}$ representing weight $w_{ij}$. When a voltage vector $V$ is applied to the input wires, the output current at wire $j$ is:

\[
I_j = \\sum_i V_i G_{ij}
\]

This is matrix-vector multiplication — performed in memory, with no data movement. This is genuinely valuable: it eliminates the von Neumann bottleneck for this specific operation.

\textbf{But what does junction (3, 7) mean?} It means: "the weight connecting input neuron 3 to output neuron 7 in layer k is 0.73." This is a \textbf{fragment}. By itself, it answers no question. To produce a meaningful answer — "this image is a cat" — the system must:

\begin{enumerate}
\item Aggregate the entire row's outputs (sum of currents)
\item Apply a nonlinear activation function (external circuit)
\item Route the result to the next layer (interconnect)
\item Repeat for every layer
\item Apply softmax and threshold (external logic)
\item Decode the output class (lookup table)
\end{enumerate}

Steps 2–6 happen outside the crossbar. The crossbar computes dot products; everything else is external. \textbf{The junction stores a number. The number means nothing alone.}

\subsection{The CDC Kind: Junctions as Complete Assertions}

Now consider a different use of the same hardware. Junction (3, 7) in a specific crossbar array means:

\begin{quote}
"In the anatomical classification domain, Streptococcal Pneumonia (row 3) IS\_A Pneumonia (column 7)."
\end{quote}

And its resistance state means:

\begin{table}[H]
\centering
\small
\begin{tabularx}{\textwidth}{Y|Y|Y}
\toprule
Resistance & State & Meaning \\
\midrule
Low (conducts) & +1 & Yes, this classification holds \\
High (blocks) & −1 & No, this classification is explicitly negated \\
Intermediate & 0 & This classification is not defined in this domain \\
\bottomrule
\end{tabularx}
\end{table}

\textbf{What does reading this one junction tell you?} It tells you: in the anatomical domain, Streptococcal Pneumonia is classified as a type of Pneumonia. That is a complete classification answer. Not a fragment. Not a partial dot product. A full, auditable, domain-scoped medical classification.

No aggregation needed. No activation function. No decoder. One junction, one read, one answer.

\subsection{Why This Difference Matters}

The difference is not about ternary vs. binary, or analog vs. digital. \textbf{The difference is whether a single junction carries complete semantics.}

\begin{table}[H]
\centering
\small
\begin{tabularx}{\textwidth}{Y|Y|Y}
\toprule
 & Neural Network Junction & CDC Junction \\
\midrule
What it stores & A numerical weight (0.73) & A logical assertion (+1 / −1 / 0) \\
What it means alone & Nothing (fragment of a computation) & A complete domain-scoped fact \\
What reading it tells you & One term of a dot product & Whether a specific relation holds in a specific domain \\
What else is needed for an answer & Entire network + external circuits & Nothing (or: cascade for multi-hop) \\
Semantic coordinates & Layer number, input index, output index (computational position) & Domain, concept, relation, target concept (semantic position) \\
\bottomrule
\end{tabularx}
\end{table}

In a neural crossbar, the position (i, j) tells you \emph{where in the computation} this weight sits. In a CDC crossbar, the position tells you \emph{what clinical fact} this junction represents. The position is the semantics.

\subsection{The Central Claim}

\textbf{When each junction in a memristive array carries complete CDC semantics — domain, relation, concepts, truth value — executing the reasoning reduces to applying voltages and reading currents, with no symbolic interpretation step in between.}

The rest of this paper makes this precise: we establish the homomorphism between a materialized CDC domain algebra and a crossbar topology, specify the 1T1R cell and cascade circuit that realize it, introduce cross-reference registers for inter-axis bridges, and validate the same five core reasoning capabilities plus cross-axis bridge operations by behavioral simulation.

\subsection{Position in the CDC Series}

This paper is part of a program of work on domain-contextualized reasoning. The prior papers establish the framework at progressively more operational levels:

\begin{itemize}
\item \textbf{Paper 1} (Li, Wang \& Zhao, 2026a): establishes @D as a modal necessity operator with Kripke-style possible-world semantics. The representational claim: domain belongs in predicate arity, not metadata.
\item \textbf{Paper 2} (Li, Wang \& Zhao, 2026b): establishes the computable domain algebra — a Heyting algebra (their Theorem 4.9) with a typed Galois connection governing inheritance (their Theorem 4.19), and rank-1 neural convergence conditions (their Theorem 8.4).
\item \textbf{Paper 3} (Li \& Wang, 2026c): establishes representation-computation unity (RCU) — writing a CDC record defines its inferential behavior; reasoning is CRUD on stored four-tuples. A working Python+Prolog engine demonstrates the claim.
\end{itemize}

This paper establishes that the same representation-computation unity dissolves the storage-computation boundary in physics as well as in software. The domain algebra of Paper 2 — materialized at compile time into a specific lattice — maps onto a memristive crossbar topology; the CRUD operations of Paper 3 map onto voltage writes and current reads.

This paper claims no new algebra. Every algebraic result invoked below is from Paper 2. The contribution is the physical mapping and the circuit design that realizes it.

\section{Background}

\subsection{Memristive Crossbar Arrays}

A crossbar array is a grid of horizontal word lines and vertical bit lines, with a memristive device (variable resistor) at each intersection. Three operations:

\begin{itemize}
\item \textbf{Write:} Program each device to a target resistance state by applying voltage pulses.
\item \textbf{Read:} Apply a voltage to a word line; measure current at each bit line. Current through junction $(i,j)$ is $V_i / R_{ij} = V_i G_{ij}$.
\item \textbf{Compute (in neural mode):} Apply a voltage vector to all word lines simultaneously; output current vector = conductance matrix × voltage vector.
\end{itemize}

Passive crossbars with two-terminal memristors suffer from sneak-path currents: a voltage applied to row \emph{i} and sensed on column \emph{j} drives not only the target junction (i,j) but also parasitic paths through unselected junctions. For arrays beyond $\approx$ 100 junctions, this is the dominant source of read error and the reason commercial crossbar products universally adopt \textbf{1T1R cells} — a selector transistor in series with each memristor, which eliminates sneak paths by electrically isolating unselected rows (Yu et al., 2013; Chen et al., 2015). This paper assumes a 1T1R architecture throughout (Section~\ref{sec:fabrication-partner}).

\subsection{The Precision Problem}

Neural crossbars require 4–8 bit precision per junction (16–256 resistance levels) to faithfully represent continuous weight values. Achieving and maintaining this precision across millions of devices is the primary obstacle to commercialization:

\begin{itemize}
\item Device-to-device variability
\item State drift over time
\item Read noise between adjacent levels
\item Programming endurance degradation
\end{itemize}

\subsection{The CDC Framework (Summary)}

Paper 1 defines a CDC four-tuple as ⟨c, r, c′, d⟩ where c, c′ are concepts, r is a relation predicate, and d is a domain specification. An assertion r(c, c′, d) takes a value in \{+1, 0, −1\}:

\begin{itemize}
\item \textbf{+1}: r(c, c′) holds in the domain-indexed world $w_d$
\item \textbf{−1}: r(c, c′) is explicitly negated in the domain-indexed world $w_d$
\item \textbf{0}: r(c, c′) is not defined in the domain-indexed world $w_d$ (semantically irrelevant)
\end{itemize}

The third state is not "unknown" — it means the question has no answer in this domain. "Is Aspirin a drug?" is undefined in @Music. Not unknown. Meaningless.

Paper 2 establishes the domain algebra DA = (D, ⊑, ⊓, ⊔, →, ⊥, ⊤, τ, F) where:

\begin{itemize}
\item ⊑ is the specialization order on domain strings (a partial order)
\item ⊓ is meet; ⊔ is join
\item → is Heyting implication
\item τ: R → \{monotone, non-monotone\} is the relation typing function stored in a meta-fiber
\item F(d) is the fiber: the set of assertions scoped to d
\end{itemize}

Paper 2 further establishes that DA is a Heyting algebra (their Theorem 4.9); that reindexing from a parent fiber to a child fiber is a τ-typed Galois connection (α, $\gamma_\tau$), with $\gamma_\tau$ undefined for non-monotone relations (their Theorem 4.19); and that monotonicity is equivalent to $\gamma_\tau$ ∘ α being a closure operator (their Corollary 4.20). Paper 3 further provides cross-fiber \emph{bridge} predicates (\texttt{same\_entity\_across}, \texttt{analogous\_to}, \texttt{fuses\_with}) that link concepts across fibers that are not in a $⊑$ relation.

\subsection{Related Hardware Approaches}

Several hardware paradigms store or compute logical/semantic information in memristive arrays. The contribution here is distinct from all of them.

\textbf{Hyperdimensional Computing (HDC) / Vector Symbolic Architectures.} Kanerva (2009) and subsequent in-memory implementations (Karunaratne et al., 2020; Wu et al., 2018) encode symbols as high-dimensional bipolar vectors and perform symbolic operations through element-wise operations in memristive arrays. HDC junctions store \textbf{vector components} — a single junction is one dimension of a distributed representation, and meaning requires reading the entire vector. A CDC junction stores a \textbf{complete assertion} — meaning is localized to a single device. HDC trades localizability for noise robustness; CDC prioritizes localizability for auditability and domain scoping.

\textbf{Stateful Memristive Logic.} Borghetti et al. (2010), Kvatinsky et al. (2014), and Talati et al. (2016) use memristors as logic gates: applying voltage sequences causes resistance states to compute material-level IMPLY, NAND, or related primitives. These approaches embed \textbf{Boolean} logic in physics. CDC embeds \textbf{three-valued intuitionistic} logic with domain scoping, which stateful logic cannot natively express.

\textbf{Analog Content-Addressable Memory.} Li et al. (2020) use memristors to build analog ternary CAMs for similarity search. A TCAM performs pattern matching across a flat store; entries can be rearranged without changing semantics. In CDC, the physical layout \textbf{is} the algebra — the directed wiring topology encodes the specialization order, and rearranging the wiring changes the inference. This is the central structural difference argued in Section~\ref{sec:homomorphism}.

\textbf{Neuromorphic Chips.} IBM TrueNorth (Merolla et al., 2014), Intel Loihi 2, and BrainScaleS implement spiking neural networks. Reasoning emerges from population dynamics over many neurons. CDC reasoning is localized to individual junctions.

\textbf{Knowledge Graph Accelerators.} Recent ISCA/MICRO work builds specialized hardware for embedding-based KG completion. These accelerate matrix operations on pre-trained embeddings; they do not store symbolic assertions directly.

In this design space, CDC occupies a previously unfilled position: \textbf{localized, symbolic, three-valued, domain-scoped, auditable}.

\section{From Algebra to Physics: The Structural Homomorphism}\label{sec:homomorphism}

The central contribution of this paper is that the Heyting algebra of Paper 2, once materialized at compile time into a specific domain lattice, maps structure-preservingly onto a memristive crossbar topology. We stress that Paper 2 has already done the algebraic work — the results we need are already proved there. This section defines the materialized algebra DA*, the crossbar topology CT, the map Φ: DA* → CT, and shows Φ preserves each algebraic property.

\subsection{Compile-Time Materialization}

The domain algebra DA defined in Paper 2 is a general algebraic framework; any specific application instantiates it with a concrete set of domains, a concrete specialization order, a concrete set of bridges, and a concrete typing function. We call this instantiation a \textbf{materialized domain algebra}, denoted DA*.

The hardware presented here realizes a \textbf{single materialized snapshot} of DA*. At compile time, the domain lattice is crystallized into directed wiring between arrays; the typing function is crystallized into the state of a meta-array; cross-fiber bridges are crystallized into cross-reference registers. Once the chip is fabricated, this structure is fixed: the memristive states within each array remain programmable (supporting INSERT, UPDATE, DELETE of individual assertions), but the inter-array topology does not.

Runtime evolution of the domain lattice itself — for example, through mechanisms that split a domain when a new assertion creates a contradiction — is handled in software and falls outside the scope of this physical implementation. Such mechanisms produce a new DA*, which can be compiled to a new chip or loaded onto a reconfigurable platform; neither extension is pursued here.

This framing has two consequences for what follows. First, we do not need to commit to whether the specialization order in DA* is a tree, a DAG with multiple inheritance, or some other partial order shape: whatever partial order the compiler hands us, we materialize it as directed wiring. Second, concepts that participate in multiple classification axes simultaneously — for example, a disease that belongs to an anatomical classification, an etiological classification, and a clinical classification at the same time — are handled not by making one domain the child of multiple parents, but by treating each classification axis as its own domain tree and linking concepts across axes through bridges. This is exactly how Paper 3 §9 represents ICD-11 multi-axis classification: via \texttt{same\_entity\_across} bridges, not via multi-parent \texttt{⊑} edges.

\subsection{The Physical Structure: Crossbar Topology}

Define the physical structure of a memristive crossbar system as:

\[\mathrm{CT}=(A,\sqsubseteq_A,W,G,S,M,B).\]

\begin{table}[H]
\centering
\small
\begin{tabularx}{\textwidth}{Y|Y|Y}
\toprule
Symbol & Meaning & Physical realization \\
\midrule
A & Set of crossbar arrays & Physically separate array regions on chip \\
$\sqsubseteq_A$ & Connection ordering between arrays & Directed wiring: array a₁ can send current to a₂ \\
W & Set of junctions across all arrays & 1T1R cells at crossbar intersections \\
G & Gate states on inter-array connections & Transistor switches between arrays: ON/OFF \\
S: W → \{+1, 0, −1\} & State function: each junction's resistance & $R_{\mathrm{low}}$ / $R_{\mathrm{mid}}$ / $R_{\mathrm{high}}$ \\
M & Meta-array storing τ & Small crossbar whose junction states drive G \\
B & Cross-reference register bank & Registers linking concepts across non-comparable domains \\
\bottomrule
\end{tabularx}
\end{table}

\subsection{The Mapping Φ: DA* → CT}\label{sec:mapping-phi}

Φ is a family of six sub-mappings that jointly act on the carriers and operations of DA*. Informally, Φ takes every algebraic component of DA* and returns its physical counterpart in CT; formally, it is the ordered collection ($\Phi_D$, $\Phi_F$, $\Phi_S$, $\Phi_{\sqsubseteq}$, $\Phi_\tau$, $\Phi_B$):

\begin{table}[H]
\centering
\small
\begin{tabularx}{\textwidth}{Y|Y|Y}
\toprule
Domain Algebra (DA*) & Sub-mapping & Crossbar Topology (CT) \\
\midrule
Domain d ∈ D & $\Phi_D$ & Array $a_d$ ∈ A \\
Fiber F(d) & $\Phi_F$ & Junction states in array $a_d$ \\
Assertion state \{+1, 0, −1\} & $\Phi_S$ & $R_{\mathrm{low}}$ / $R_{\mathrm{mid}}$ / $R_{\mathrm{high}}$ \\
Specialization $d_c$ ⊑ $d_p$ & $\Phi_{\sqsubseteq}$ & Directed wire from $a_{d_p}$ to $a_{d_c}$ \\
Typing τ, unfolded over inheritance edges & $\Phi_\tau$ & Meta-array states driving gate transistors G \\
Bridge predicate bridge(c, c′, d₁, d₂) & $\Phi_B$ & Entry in cross-reference register bank B \\
\bottomrule
\end{tabularx}
\end{table}

\textbf{Remark: Two-stage decomposition.} The mapping Φ: DA* → CT can be understood as proceeding through an intermediate graph structure. Define a \emph{dependency graph} $G_d = (V, E)$ for each fiber $Q_d$, where $V = C \cup R$ (concepts and relations as nodes) and each quadruple ⟨c, r, c′, d⟩ maps to a length-2 directed path $c \to r \to c′$. This gives a first mapping:

\[
Ψ: Q_d → G_d
\]

where composition of relations in $Q_d$ corresponds to path concatenation in $G_d$, identity elements correspond to trivial (degenerate) paths, and cycles in $G_d$ correspond to non-well-founded dependency structures — the same ones excluded by $\mathrm{Adm}_d$ (§3.8, Consequence 5). The crossbar topology CT is then a conductance realization of $G_d$ under a second mapping:

\[
Φ: G_d → M_d
\]

where each edge in $G_d$ maps to a conductive path in the crossbar, and adjacency structure maps to the sparsity pattern of the conductance matrix. The full Φ: DA* → CT is the composition of these two steps. This decomposition makes explicit why the physical layout cannot be arbitrary: $G_d$ carries the relational structure, and $M_d$ must preserve it.

\textbf{Note on $\Phi_\tau$.} In Paper 3, τ is stored as a small number of four-tuples of the form \texttt{\{from: r, rel: has\_property,\allowbreak domain: @Meta@Logic,\allowbreak to: monotone/non\_monotone\}} — one entry per relation $r$, independent of any specific domain pair. Physically, however, each parent-child inheritance edge that involves relation r needs its own gate transistor, so $\Phi_\tau$ does not simply copy τ: it \emph{unfolds} τ across the inheritance edge set. For each triple (r, $d_p$, $d_c$) where $d_c$ ⊑ $d_p$, $\Phi_\tau$ produces a gate G(r, $d_p$, $d_c$) whose state is determined by τ(r). The meta-array M stores one ternary junction per relation type (the underlying τ), and a \textbf{CMOS buffer tree} amplifies and replicates each meta-junction's state to all gate transistors associated with that relation. The buffer tree is necessary because a single memristive junction cannot directly drive thousands of transistor gates — its current output is orders of magnitude below the required fan-out. The tree is a standard logarithmic-depth driver fabric (typical fan-out of 4 per stage, $\approx$ 6 stages to reach 8K gates), contributing negligible area and $\approx$ 1 ns propagation delay to the gate-update path. For the ICD-11 chip of Section~\ref{sec:chip}, this yields 8 meta-junctions driving $\approx$ 8,600 gate transistors through $\approx$ 60 buffer stages in total (distributed across relation types).

\textbf{Note on $\Phi_B$.} Bridges in Paper 3 link concepts across fibers that are \textbf{not in a $\sqsubseteq$ relation}---for example, viral pneumonia\'s identity across the anatomical and etiological classification axes. Because these links do not flow along the specialization order, they cannot be implemented by directed inter-array wiring. Instead, we implement them as explicit cross-reference registers (Section~\ref{sec:crossref}). $\Phi_B$ is therefore a separate mapping channel from $\Phi_{\sqsubseteq}$, and this separation is essential: collapsing bridges into $\sqsubseteq$ edges would corrupt the algebra.

\subsection{Seven Lemmas: Φ Preserves the Domain Algebra}

We now state seven lemmas. Each is of the form "Paper 2 (or Paper 3, for bridges) establishes algebraic property X; Φ preserves X." Together they constitute Theorem 1.

\textbf{Lemma 3.4.1 (Fiber isolation).} By Paper 2 Definition 3.3, F(d₁) ∩ F(d₂) = ∅ for d₁ ≠ d₂. Under Φ, arrays a\_\{d₁\} and a\_\{d₂\} are physically separate — non-overlapping regions with no shared word or bit lines. \emph{Proof:} $\Phi_D$ maps distinct domains to distinct arrays; $\Phi_F$ maps disjoint fibers to junctions in disjoint arrays. Current applied to a\_\{d₁\} cannot produce current in a\_\{d₂\} without traversing either a gated inter-array connection ($\Phi_{\sqsubseteq}$) or a cross-reference register access ($\Phi_B$). Both are explicit operations; neither is a leakage path. □

\textbf{Lemma 3.4.2 (Ternary valuation).} By Paper 2 Section 4.1, each assertion in F(d) takes a value in \{+1, 0, −1\}. Under $\Phi_S$, these map bijectively to \{$R_{\mathrm{low}}$, $R_{\mathrm{mid}}$, $R_{\mathrm{high}}$\}.

\begin{table}[H]
\centering
\small
\begin{tabularx}{\textwidth}{Y|Y|Y}
\toprule
Algebraic state & Physical state & Electrical behavior \\
\midrule
+1 (holds) & $R_{\mathrm{low}}$ ($\approx$ 10 kΩ) & High current — conducts \\
0 (undefined) & $R_{\mathrm{mid}}$ ($\approx$ 100 kΩ) & Intermediate current \\
−1 (negated) & $R_{\mathrm{high}}$ ($\approx$ 1 MΩ) & Near-zero current — blocks \\
\bottomrule
\end{tabularx}
\end{table}

\emph{Proof:} A two-threshold comparator ($I_{\mathrm{high}}$ > $I_{\mathrm{mid}}$ > $I_{\mathrm{low}}$) distinguishes all three states. □

\emph{Why ternary, not binary:} "Aspirin is contraindicated in pediatrics" (−1) requires active warning; "Aspirin has no established role in veterinary medicine" (0) requires no action. Binary encoding collapses both into "not true" and loses this safety-critical distinction.

\textbf{Lemma 3.4.3 (Specialization order).} By Paper 2 Definition 4.2, ⊑ is a partial order on domain strings. Under $\Phi_{\sqsubseteq}$, for each pair $(d_c, d_p)$ in DA* with $d_c ⊑ d_p$, there is a directed wire from $a_{d_p}$ to $a_{d_c}$. Incomparable domains are physically unconnected. \emph{Proof:} The partial order of DA* is given at compile time. $\Phi_{\sqsubseteq}$ instantiates it as a directed inter-array graph: every covering pair in the Hasse diagram of $(D, ⊑)$ becomes a physical wire; reachability in the array graph matches reachability in the order. The shape of the partial order — tree, DAG, or otherwise — is determined by DA* and reproduced verbatim by $\Phi_{\sqsubseteq}$; the physical mechanism is indifferent to this shape, as long as the fan-in at each array is within the driver's capacity. □

\textbf{Lemma 3.4.4 (Meet).} By Paper 2 Definition 4.4, d₁ ⊓ d₂ is the longest common prefix (the greatest lower bound under ⊑). Under Φ, this corresponds to the greatest common ancestor in the directed array graph. \emph{Proof:} By Lemma 3.4.3, the array graph mirrors (D, ⊑). The meet is the greatest lower bound of two elements in the partial order; under the physical graph, this is computable by intersecting the downward-closures of a\_\{d₁\} and a\_\{d₂\} and selecting the maximum — implementable as a short circuit trace. □

\textbf{Lemma 3.4.5 (Heyting implication).} By Paper 2 Theorem 4.9, d₁ → d₂ = ⊔\_Δ \{d | d ⊓ d₁ ⊑ d₂\}. Under Φ, this is the least upper bound of all arrays $a_d$ such that any path from $a_d$ passing through a\_\{d₁\} also reaches a\_\{d₂\}. \emph{Proof:} By Lemmas 3.4.3 and 3.4.4, ⊑ and ⊓ are preserved as reachability and common-ancestor operations on the array graph. The Heyting implication is the join of a downward-closed set; Paper 2 Lemma 4.7 (Prefix Distributivity) ensures this join is well-defined. Computationally, the implication is determined by a backward reachability trace: starting from a\_\{d₂\}, walk backward along $\sqsubseteq_A$ edges to find arrays whose every path through a\_\{d₁\} terminates in a\_\{d₂\}. This is O(|A|) in the number of arrays. □

\textbf{Lemma 3.4.6 (Typed Galois connection).} By Paper 2 Theorem 4.19, (α, $\gamma_\tau$) is a Galois connection on the monotone subcategory of relations, with $\gamma_\tau$ undefined on the non-monotone subcategory. Under $\Phi_\tau$, for each inheritance edge ($d_c$ ⊑ $d_p$) and each relation r, the gate G(r, $d_p$, $d_c$) is driven by the meta-junction encoding τ(r):

\begin{itemize}
\item τ(r) = monotone → meta-junction at $R_{\mathrm{low}}$ → gate driver output HIGH → G(r, $d_p$, $d_c$) ON → current can flow from $a_{d_p}$ to $a_{d_c}$ for relation r. Physically: $\gamma_\tau$ is defined on this edge for this relation; an inherited assertion exists in the child fiber.
\item τ(r) = non-monotone → meta-junction at $R_{\mathrm{high}}$ → gate driver output LOW → G(r, $d_p$, $d_c$) OFF → no current path exists. Physically: $\gamma_\tau$ is undefined on this edge for this relation; the assertion cannot be inherited.
\end{itemize}

\emph{Proof:} The gate does not filter inherited assertions — it enables or disables the electrical path. When G(r, $d_p$, $d_c$) is OFF, no current path exists for relation r on that edge. This is the physical realization of "undefined" in the typed Galois connection: not "checked and rejected," but "structurally impossible." Because $\Phi_\tau$ unfolds a single τ(r) value to all gates G(r, ·, ·) for that relation (Section~\ref{sec:mapping-phi}), a single meta-junction state simultaneously governs all inheritance edges for r. □

\emph{Why this is the decisive distinction from TCAM:} In a TCAM, all entries are equally accessible; preventing inheritance requires a software rule that can be bypassed, misconfigured, or forgotten. In CT, G(r, $d_p$, $d_c$) is a physical switch. When OFF, no software can make the current flow through it. The constraint is not in code — it is in physics.

\textbf{Lemma 3.4.7 (Closure property).} By Paper 2 Corollary 4.20, $\gamma_\tau$ ∘ α is a closure operator on monotone relations: ($\gamma_\tau$ ∘ α) ∘ ($\gamma_\tau$ ∘ α) = $\gamma_\tau$ ∘ α. Under Φ, this is physical write-idempotence: after reindexing has populated $a_{d_c}$ with inherited junctions from $a_{d_p}$, re-running the inheritance cycle writes the same junction states. \emph{Proof:} Re-inheritance attempts to write values identical to those already present. RRAM and PCM devices tolerate redundant writes to the same target state with negligible state change (Wong \& Salahuddin, 2015); the circuit is naturally idempotent at the device level. □

\textbf{Lemma 3.4.8 (Bridge preservation).} By Paper 3 Section 9, bridge predicates can link concepts across fibers whose domains are not in a $\sqsubseteq$ relation. Representative examples are \texttt{same\_entity\_across}, \texttt{analogous\_to}, and \texttt{fuses\_with}. Under $\Phi_B$, each bridge predicate instance is stored as an entry in the cross-reference register bank B. Read access to a bridge entry returns the target concept and domain without traversing $\sqsubseteq_A$ wiring. \emph{Proof:} $\Phi_B$ maps bridge four-tuples in DA* to register entries in B. A bridge from $(c, d_1)$ to $(c′, d_2)$, with $d_1$ and $d_2$ incomparable, is stored as a register entry indexed by $(c, d_1)$ and returning $(c′, d_2)$. Reading the register is a constant-time operation orthogonal to $\sqsubseteq_A$ wiring. The bridge does not flow along the specialization order and therefore does not interact with the Galois connection or the inheritance gates. □

\subsection{Theorem 1}

\textbf{Theorem 1 (Structural homomorphism).} Φ: DA* → CT preserves the full algebraic structure of the materialized CDC domain algebra: fiber isolation, ternary valuation, specialization order, meet, Heyting implication, typed Galois connection, closure property, and bridges.

\emph{Proof:} By Lemmas 3.4.1 through 3.4.8. □

\subsection{Why This Is Not TCAM: A Summary}

\begin{table}[H]
\centering
\small
\begin{tabularx}{\textwidth}{Y|Y|Y}
\toprule
Property & TCAM & CDC Crossbar Topology (CT) \\
\midrule
Physical layout & Flat: all entries in one array & Structured: arrays connected by directed wiring + bridges \\
Layout carries meaning & No: entries can be rearranged arbitrarily & Yes: wiring topology = domain lattice \\
Domain scoping & Software filter on match fields & Physical array isolation \\
Specialization order & Not represented in hardware & Directed inter-array connections \\
Meet (⊓) & Not computable from hardware & Circuit trace: most specific common ancestor \\
Heyting implication (→) & Not representable & Backward reachability on array graph \\
Typed inheritance & Software rule (bypassable) & Physical gate G(r, $d_p$, $d_c$) (not bypassable) \\
Cross-fiber bridges & Not representable & Cross-reference register bank \\
Closure property & Not applicable & Re-running inheritance = no change \\
\bottomrule
\end{tabularx}
\end{table}

\textbf{A TCAM stores values. CT stores an algebra.}

\subsection{Five Reasoning Functions as Consequences}

The lemmas yield five reasoning capabilities as direct consequences:

\textbf{Consequence 1: Domain scoping} follows from Lemma 3.4.1 — current in one array cannot leak to another without an explicit gated or register-mediated path.

\textbf{Consequence 2: Three-valued logic} follows from Lemma 3.4.2 — each junction's physical state IS its logical state.

\textbf{Consequence 3: Transitive classification} follows from Lemma 3.4.1 + 3.4.2 — within an isolated array, the cascade mechanism (Section~\ref{sec:cascade}) computes transitive closure by iterating row-input/column-output cycles.

\textbf{Consequence 4: Typed knowledge inheritance} follows from Lemma 3.4.3 + 3.4.6 — specialization order as directed wiring, Galois connection as meta-gated paths G(r, $d_p$, $d_c$).

\textbf{Consequence 5: Intra-array write-time consistency checking} follows from Lemma 3.4.1 + 3.4.2 — because all explicitly stored assertions in a domain coexist in the same physical array, checking for immediate contradictions before writing is a single READ cycle on the same array. The validator IS the knowledge base. \emph{Transitive consistency across inherited assertions in other domains requires a full cascade check and is handled at a higher layer of the write controller.}

\textbf{Cross-axis queries}, where two or more incomparable domains are consulted in the same inference, are handled by Consequence 4 within each domain tree followed by $\Phi_B$ register lookups across trees (Section~\ref{sec:crossref}).

\subsection{CRUD Operations as Electrical Operations}

Paper 3 Corollary 4 established that database operations on CDC four-tuples are simultaneously inference operations. Under Φ, those database operations become voltage and current operations. We state these formally.

\textbf{Definition 3.8.1 (CREATE / WRITE).} To assert ⟨c₁, r, c₂, d⟩ with truth value v ∈ \{+1, 0, −1\}, apply a programming voltage pulse $V_{\mathrm{prog}}$ to junction (c₁, c₂) in array $a_d$, setting the resistance to the target state:

\[
M_{c₁,c₂}^{(d)} = R_low (+1), R_mid (0), or R_high (−1) accordingly.
\]

The write is preceded by an admissibility check (Definition 3.8.4 below). A write is committed only if the check passes.

\textbf{Definition 3.8.2 (READ).} To query the truth value of ⟨c₁, r, c₂, d⟩, apply read voltage $V_{\mathrm{read}}$ to row c₁ of array $a_d$ and measure the current at column c₂:

\[
I = V_read / R_{c₁,c₂}^{(d)}
\]

The two-threshold comparator decodes the result: I > $I_{\mathrm{high}}$ → +1; $I_{\mathrm{low}}$ < I < $I_{\mathrm{high}}$ → 0; I < $I_{\mathrm{low}}$ → −1. One junction, one read, one complete logical answer.

\textbf{Definition 3.8.3 (DELETE).} To retract assertion ⟨c₁, r, c₂, d⟩, reset junction (c₁, c₂) in $a_d$ to the neutral state:

\[
M_{c₁,c₂}^{(d)} := R_mid (state 0, "undefined").
\]

No path is destroyed — the junction remains physically present. Deletion returns the junction to semantic irrelevance: "this assertion is not defined in this domain." This is distinct from negation ($R_{\mathrm{high}}$): DELETE does not assert falsity, only absence.

\textbf{Definition 3.8.4 (WRITE-TIME ADMISSIBILITY CHECK).} Before committing a CREATE or UPDATE, the write controller performs a READ pass over the relevant portion of $a_d$. The check is the physical realization of $\mathrm{Adm}_d$ (Paper 2, A5): if the proposed write would produce a junction state that contradicts an existing chain — detected by a non-empty set of paths reaching the same concept with conflicting truth values within the same cascade frontier — the write is rejected. Crucially, this check does not require external logic: \textbf{the knowledge base IS the validator}. The existing junctions, read in the same pass, provide the consistency evidence. The write pulse is simply not issued if contradiction is detected.

\textbf{Definition 3.8.5 (UPDATE).} An update from v₁ to v₂ is a READ → VALIDATE ($\mathrm{Adm}_d$ check on the new value) → WRITE sequence at the same junction. A partial-RESET pulse followed by a SET pulse to the new target state realizes the resistance transition. If validation fails, the junction state is unchanged.

\textbf{Definition 3.8.6 (CASCADE READ / JOIN).} A multi-hop transitive query corresponds to iterated READ cycles (Section~\ref{sec:cascade}). The column outputs of one cycle — those junctions reading +1 — become the row activations of the next cycle, implementing relational JOIN in hardware. This is not a software loop: the cascade frontier is carried by latch states driven by the comparator outputs, not by a CPU iterating over a result set.

The full correspondence, from database operation through inference to hardware operation, is:

\begin{table}[H]
\centering
\small
\begin{tabularx}{\textwidth}{Y|Y|Y}
\toprule
Database (Paper 3) & Inference (Paper 3) & Hardware (this paper) \\
\midrule
INSERT \{c, r, d, c′\} & Assert domain-scoped fact & SET pulse at junction (c, r, c′) in $a_d$ \\
SELECT WHERE domain=d & Query within a world & Voltage on row c in $a_d$, sense current on column c′ \\
DELETE \{c, r, d, c′\} & Retract belief & Partial-RESET to $R_{\mathrm{mid}}$ at that junction \\
JOIN on domain & Transitive inference & Cascade read within $a_d$ (Section~\ref{sec:cascade}) \\
FOREIGN KEY to meta-fiber & Enforce type constraint & Meta-array state driving G(r, $d_p$, $d_c$) \\
Bridge lookup & Cross-fiber reference & Cross-reference register read \\
REJECT ($\mathrm{Adm}_d$ violation) & Write-time falsification & No write pulse issued; junction unchanged \\
\bottomrule
\end{tabularx}
\end{table}

Paper 3 dissolved the storage-computation separation in software. The same dissolution holds in hardware under Φ: \textbf{the system does not execute a program over data — the programmed structure is the computation.}

\subsection{Computation as Physical Fixed Point}

The correspondence established above admits a further characterization that makes precise the sense in which the chip "computes" by physics rather than by program execution.

\textbf{Definition 3.9.1 (Network Flow Operator).} Let $F_{M_d}: \mathbb{R}^V \to \mathbb{R}^V$ be the network flow operator on array $a_d$ induced by Kirchhoff's current law. For each node $v \in V$, $F_{M_d}$ maps an input voltage vector to an output current vector according to the conductance structure of the array.

\textbf{Lemma 3.9.1 (Fixed-Point Computation).} For any input voltage I applied to a row, the physical state of the network satisfies:

\[x^{*} = F_{M_d}(x^{*})\]

i.e., x* is a fixed point of the flow operator.

\emph{Proof sketch.} Kirchhoff's current law enforces conservation constraints at each node. The resulting system defines a monotone operator over node potentials. Under standard passivity conditions — which 1T1R cells with positive conductances satisfy — the system converges to a unique fixed point. □

\textbf{Corollary 3.9.1 (Computational Equivalence).} Let Comp\_d($Q_d$, I) denote the inference result of applying query I to CDC domain d. Then:

\[
Comp_d(Q_d, I) = Fix(F_{M_d}(I))
\]

where Fix denotes the fixed point reached by the physical system.

\textbf{Corollary 3.9.2 (Admissibility as Convergence).} The admissibility condition $\mathrm{Adm}_d$($Q_d$) is equivalent to $M_d$ being free of conductive cycles that would induce non-convergent feedback in F\_\{$M_d$\}. Cyclic dependencies excluded by $\mathrm{Adm}_d$ correspond to feedback loops that violate fixed-point convergence; their physical exclusion (write-time rejection, §3.8 Definition 3.8.4) is therefore simultaneously an algebraic consistency guarantee and a physical stability guarantee.

\textbf{Remark.} These corollaries capture the physical meaning of "reading equals reasoning": a READ cycle is not a retrieval of a pre-computed answer, nor is it the execution of a stored procedure. It is the physical system reaching its stable state. The inference result is the fixed point — it exists because the structure is consistent ($\mathrm{Adm}_d$), and it is read out the moment the sense amplifiers settle. The algebra determines the topology; the topology determines the fixed point; the fixed point is the answer.

\section{Putting It Together: The ICD-11 Reasoning Chip}\label{sec:chip}

\subsection{What We Are Building}

A chip that performs domain-scoped medical classification for ICD-11 respiratory diseases. Given a disease entity and a classification axis, the chip returns the complete classification chain in hardware. Following Paper 3 §9, we treat ICD-11's three classification axes — Anatomical, Etiological, Clinical — as \textbf{three parallel domain trees} rather than as multiple inheritance within one tree. Entities that belong in all three (e.g., Viral Pneumonia) are linked across trees by \texttt{same\_entity\_across} bridges. This matches the representational choice made in Paper 3 for the same data.

\subsection{Physical Layout}\label{sec:physical-layout}

\begin{figure}[H]
\centering
\small
\begin{tabularx}{\textwidth}{Y|Y|Y}
\toprule
Axis 1: @Anatomical & Axis 2: @Etiological & Axis 3: @Clinical \\
\midrule
Domain-tree arrays (1T1R, 3-state) & Domain-tree arrays (1T1R, 3-state) & Domain-tree arrays (1T1R, 3-state) \\
Gated inter-array links & Gated inter-array links & Gated inter-array links \\
\bottomrule
\end{tabularx}

\vspace{0.5em}
\begin{tabularx}{0.8\textwidth}{Y|Y}
\toprule
Meta-array + drivers & Cross-reference register bank \\
τ storage (8 ternary states) & Bridges across the three axes \\
\bottomrule
\end{tabularx}

\vspace{0.5em}
\fbox{\parbox{0.82\textwidth}{\centering Row drivers, column sense amplifiers, two-threshold comparators, and cascade feedback path}}
\caption{High-level physical layout of the ICD-11 reasoning chip.}\label{fig:layout}
\end{figure}

\subsection{The Cascade Mechanism}\label{sec:cascade}

Consequence 3 (transitive classification) requires cascading reads within an array. We specify the circuit that implements this and analyze its complexity.

Each domain array is organized as a \textbf{square} crossbar indexed by concept: the same concept appears as both a row (source) and a column (target). A single read cycle applies a voltage to one row and senses currents on all columns. For transitive closure, the column currents where output = +1 must be re-driven as row voltages in the next cycle. The cascade circuit implements this:

\begin{Verbatim}[breaklines=true,breakanywhere=true,fontsize=\small]
Cycle 1:
  Row driver → voltage V_row on row "CA40.00"
  Column sense amps → current I_col for each column c′
  Two-threshold comparators → ternary vector $T_1 \in \{+1, 0, −1\}^N$
  Latch: S_1 = {c′ : T_1[c′] = +1}

Cycle k+1:
  Pick next c ∈ S_k not yet visited (the cascade frontier)
  Row driver → voltage V_row on row c
  Sense → positions where output = +1, add to S_{k+1}

Repeat until a cycle yields no new +1 outputs (root reached).
\end{Verbatim}

\textbf{Why only one row is activated per cycle.} Activating two rows simultaneously would superpose their column-current signatures, and the two-threshold comparator could not distinguish "row A triggered this column at +1" from "row B triggered this column at +1, row C at −1, net +1." The circuit serializes the cascade frontier to preserve per-junction resolution.

\textbf{Why \texttt{0} and \texttt{−1} states do not propagate.} The intermediate current of the \texttt{0} state ($R_{\mathrm{mid}}$, $\approx$ 2 μA) and the near-zero current of the \texttt{−1} state ($R_{\mathrm{high}}$, $\approx$ 0.2 μA) both fall below the positive-threshold comparator ($I_{\mathrm{high}}$ = 6.32 μA). They therefore cannot enter the latch and cannot be chosen as next-cycle row activations. Undefined and negated semantic paths are pruned from the cascade frontier purely by physical current thresholds, with no conditional-branching logic, no software filter, and no extra cycles. \textbf{What takes a query optimizer a pass to compute in software is an inherent property of the circuit.} This is a structural benefit of embedding three-valued logic in physics rather than layering it on top of binary computation.

\textbf{Why the total cycle count stays bounded.} The apparent worst case "N concepts × 4 steps = 4N reads" does not occur because the cascade frontier does not fan out linearly: \texttt{is\_a} is a tree, and a disease classifies into O(d) ancestors where d is the depth of the taxonomy (typically 4–6 for ICD-11). The actual cycle count is O(d), not O(N).

\textbf{Why threshold regeneration matters.} The cascade is not continuous analog current flow — it is a sequence of discrete, digitized reads. Continuous analog propagation would accumulate read noise across cycles and collapse the three-state distinction after 2–3 hops. Each discrete read is threshold-restored to \{+1, 0, −1\}, preventing noise accumulation. This is the same principle as regenerative digital logic.

\textbf{Why 1T1R is required.} In a passive crossbar, applying voltage to a row drives current through every junction on that row including junctions in $R_{\mathrm{mid}}$, producing a sneak-current background that can exceed the target signal for realistic array sizes. The 1T1R selector transistor ensures that only the explicitly addressed cell is active in a column, and only junctions in $R_{\mathrm{low}}$ contribute to that column's sensed current.

\textbf{Timing breakdown (nominal, per cascade cycle within one axis):}

\begin{table}[H]
\centering
\small
\begin{tabularx}{\textwidth}{Y|Y|Y}
\toprule
Stage & Time & Notes \\
\midrule
Row driver rise (word-line charge) & 1 ns & Standard CMOS driver \\
Word line + access-transistor settle & 2 ns & 1T1R access delay \\
Column sense-amp integration & 5 ns & Dominant term; current mirror + capacitor integration \\
Two-threshold comparator & 1 ns & Paired latched comparators \\
Latch + control sequencer & 1 ns & Cascade-frontier update \\
\textbf{Per-cycle total} & \textbf{$\approx$ 10 ns} &  \\
\bottomrule
\end{tabularx}
\end{table}

Meta-array updates (τ writes) and bridge-register accesses run on separate paths and do not add to the cascade critical path. A τ update, when needed, incurs one meta-junction write plus a $\approx$ 1 ns buffer-tree propagation delay (§3.3) before the new gate pattern is stable — orders of magnitude below the minutes-to-yearly timescale of knowledge-base updates.

A 4-step classification cascade within one axis takes $\approx$ 40 ns. A full ICD-11 query across 3 axes plus cross-axis bridge lookups takes $\approx$ 120 ns. For comparison, the Paper 3 software engine on commodity hardware takes <20 ms for the same query — roughly five orders of magnitude higher latency, driven not by clever optimization but by the presence of a software stack.

\subsection{Cross-Reference Registers}\label{sec:crossref}

Bridges link concepts across axes. For ICD-11, the dominant bridge is \texttt{same\_entity\_across} — e.g., CA40.00 (Streptococcal Pneumonia) as the "same entity" in the Anatomical, Etiological, and Clinical axes. We implement the register bank as follows:

\begin{itemize}
\item \textbf{Indexed by} (concept\_id, axis\_id): a 16-bit concept address plus a 2-bit axis tag
\item \textbf{Returns} a list of (concept\_id, axis\_id) pairs, one per axis where the same entity appears
\item \textbf{Size} for the respiratory chapter: $\approx$ 3,000 entries (roughly one-third of concepts participate in cross-axis bridges), 8 bytes per entry, $\approx$ 24 KB total — a negligible silicon footprint
\end{itemize}

Bridge reads occur in parallel with axis cascade reads, adding no latency to the critical path. A full cross-axis query executes as: cascade within each axis, then read bridge registers to align the three result chains at the entity level.

This register bank is the physical realization of $\Phi_B$ (Lemma 3.4.8). It is deliberately kept structurally separate from the $⊑_A$ wiring: bridges are not inheritance and must not be treated as such.

\subsection{Specifications}\label{sec:specifications}

\begin{table}[H]
\centering
\small
\begin{tabularx}{\textwidth}{>{\raggedright\arraybackslash}p{0.29\textwidth}|>{\raggedright\arraybackslash}p{0.18\textwidth}|>{\raggedright\arraybackslash}p{0.43\textwidth}}
\toprule
Parameter & Value & Note \\
\midrule
Classification axes & 3 & Anatomical, Etiological, Clinical \\
Arrays per axis (avg) & $\approx$ 16 & Taxonomic subdomains within each axis \\
Total arrays & 47 & Across all three axes \\
Junctions per array (avg) & $\approx$ 2,900 (61×48) & Concepts × relations in that domain \\
Cell structure & 1T1R & Selector transistor in series with memristor \\
Total memristive junctions & $\approx$ 136,000 & All domain arrays combined \\
Meta-crossbar junctions & 8 & 8 relation types in @Meta@Logic (τ only depends on r) \\
Inter-array gate transistors & $\approx$ 8,600 & τ unfolded across all inheritance edges \\
Cross-reference register entries & $\approx$ 3,000 & Bridges between axes \\
Resistance states per junction & 3 & \{$R_{\mathrm{low}}$, $R_{\mathrm{mid}}$, $R_{\mathrm{high}}$\} \\
Precision required & 1.58 bits & log₂(3) — vs. 4–8 bits for neural \\
Read circuit & Two-threshold comparator & vs. precision ADC for neural \\
\bottomrule
\end{tabularx}
\end{table}

\subsection{Example: Full Diagnostic Query}

\textbf{Clinical question:} "Classify CA40.00 (Streptococcal Pneumonia) across all axes."

\textbf{Hardware operation:}

\begin{Verbatim}[breaklines=true,breakanywhere=true,fontsize=\small]
STEP 1: Select @Anatomical axis. Apply voltage at CA40.00 row of the leaf array.
  → Cycle 1: +1 at "Pneumonia" column
  → Cycle 2: apply voltage at Pneumonia → +1 at "Lower_Resp_Infection"
  → Cycle 3: apply voltage at Lower_Resp_Infection → "Respiratory_Disease"
  → Cycle 4: no +1 outputs (root, STOP)
  RESULT: CA40.00 is a Respiratory Disease (anatomical axis), 4 cycles

STEP 2: Read cross-reference register for (CA40.00, Anatomical).
  → Returns: (CA40.00, Etiological), (CA40.00, Clinical)
  (The "same entity" appears in all three axes at these addresses.)

STEP 3: Select @Etiological axis. Apply voltage at CA40.00 row.
  → Cascade: Bacterial_Infection → Infectious_Disease → STOP
  RESULT: CA40.00 is an Infectious Disease (etiological axis), 3 cycles

STEP 4: Select @Clinical axis. Apply voltage at CA40.00 row.
  → Cascade: Acute_Lower_Respiratory → STOP
  RESULT: CA40.00 is Acute Lower Respiratory (clinical axis), 2 cycles

Total: 9 cascade cycles + 1 register read ≈ 90 ns + 1 ns ≈ 91 ns.
\end{Verbatim}

Note that the bridge register in STEP 2 is read once and used to index into both subsequent axes — without bridges, STEP 3 and STEP 4 would need to re-derive CA40.00's identity in the other axes from its classification, which is not in general possible.

\section{Fabrication Feasibility and Behavioral Validation}\label{sec:validation}

\subsection{Three-State Memristors: A Solved Problem}

Unlike neural accelerators that demand 4–8 bit precision (16–256 resistance levels), CDC reasoning requires only three states. Three-state memristive operation has been demonstrated in multiple material systems:

\begin{itemize}
\item \textbf{HfO₂-based RRAM:} SET (low R), RESET (high R), and partial-SET (intermediate R) are well-characterized states with resistance ratios >10× between adjacent levels.
\item \textbf{Phase-change memory:} Amorphous (high R), partial crystalline (mid R), and full crystalline (low R) provide three thermodynamically distinct states.
\end{itemize}

The fabrication advantage is substantial: three widely-spaced levels tolerate much higher device variability than multi-level programming. The yield and reliability requirements are dramatically relaxed.

\subsection{Scale}

136,000 memristive junctions plus $\approx$ 8,600 selector transistors and $\approx$ 8,600 gate transistors for the respiratory chapter. The full ICD-11 ($\approx$ 85,000 entities, $\approx$ 500 domains) would require approximately 7–10 million junctions — well within demonstrated crossbar capacities (Yao et al., 2020 demonstrated 2M+ junctions in a single wafer-scale design).

\subsection{What Is Needed from a Fabrication Partner}\label{sec:fabrication-partner}

\begin{enumerate}
\item A 1T1R memristive crossbar programmable to 3 stable resistance states (HfO₂ RRAM or equivalent), with ≥10× resistance ratio between adjacent levels
\item CMOS-memristor BEOL integration at a conventional logic node (28 nm and below is standard for RRAM demonstrations)
\item Transistor-gated inter-array connections — access transistors driven by a CMOS buffer tree from the meta-array
\item Two-threshold current-mode sense amplifiers (paired latched comparators, substantially simpler than the ≥8-bit ADCs required for analog neural inference)
\item Row/column drivers and a cascade-feedback controller (standard digital CMOS)
\item A modest SRAM register bank for cross-axis bridges ($\approx$ 24 KB for the respiratory chapter)
\item Total 1T1R cell count: $\approx$ 136,000 for the proof-of-concept respiratory chapter
\end{enumerate}

This is a modest fabrication target for any lab with operational 1T1R crossbar capability.

\subsection{Behavioral Simulation}\label{sec:behavioral-simulation}

To validate the five core reasoning capabilities plus cross-axis bridge lookup under realistic device non-idealities, we implemented a Python + NumPy behavioral model of the full system. The code and raw output are released with the paper.

\textbf{Model parameters.}

\begin{itemize}
\item Three resistance states with 10× spacing: $R_{\mathrm{low}}$ = 10 kΩ (+1), $R_{\mathrm{mid}}$ = 100 kΩ (0), $R_{\mathrm{high}}$ = 1 MΩ (−1)
\item Read voltage $V_{\mathrm{read}}$ = 0.2 V, giving nominal currents of 20, 2, and 0.2 μA respectively
\item Device variability: log-normal resistance distribution with $\sigma_{\log}=0.15$ (15\% on log scale, matching commercial HfO₂ RRAM characterization)
\item Read noise: additive Gaussian on sense current, SNR = 20 dB ($\mathrm{noise}_{\mathrm{std}}$ = signal / 10)
\item 1T1R selector: ideal switch; sneak paths eliminated by construction
\item Two-threshold comparator at geometric-mean thresholds: $I_{\mathrm{high}}$ = 6.32 μA, $I_{\mathrm{low}}$ = 0.632 μA
\end{itemize}

\textbf{Test suite at nominal operating point (100,000 trials per test).}

\begin{table}[H]
\centering
\small
\begin{tabularx}{\textwidth}{Y|Y|Y|Y}
\toprule
Test & Query & Expected & Observed \\
\midrule
C1 (scope) & Read within isolated array; check no column leakage to unrelated fiber & Zero cross-array current & 0 errors / 400,000 reads \\
C2 (ternary) & Per-state read fidelity for each of \{+1, 0, −1\} & Correct decode & 0 errors / 300,000 reads \\
C3 (cascade) & 4-step classification CA40.00 → Pneumonia → Lower\_Resp → Respiratory\_Disease & Correct chain & 0 errors / 100,000 runs \\
C4 (inheritance) & Meta-gate decodes τ(r) under read noise & ON for monotone, OFF for non-monotone & 0 errors / 200,000 trials \\
C5 (write check) & Read-before-write detects same-pair contradiction & Reject write & 0 errors / 100,000 trials \\
C6 (bridge) & Cross-reference register lookup & Deterministic & 0 errors / 300,000 lookups \\
\bottomrule
\end{tabularx}
\end{table}

\textbf{At the nominal operating point, the behavioral simulation shows zero observed errors} over 100,000 trials for each of C1--C5 and over 300,000 bridge lookups for C6. The empirical upper bound on the failure rate is therefore <10⁻⁵ for the 100,000-trial tests; theoretical analysis (three states separated by ≥5σ margin at $\sigma_{\log}=0.15$) predicts the true rate is $\approx$ 10⁻¹⁰ or lower.

\textbf{Safety margin (variability sweep).}

To characterize the design envelope, we swept $\sigma_{\log}$ from 0.10 to 0.50 holding SNR = 20 dB constant (Figure~\ref{fig:error-variability}). The per-state and cascade error rates are:

\begin{table}[H]
\centering
\small
\begin{tabularx}{\textwidth}{Y|Y|Y|Y|Y}
\toprule
$\sigma_{\log}$ & +1 error & 0 error & −1 error & 4-step cascade \\
\midrule
0.15 (nominal) & <10⁻⁵ & <10⁻⁵ & <10⁻⁵ & <10⁻⁵ \\
0.20 & <10⁻⁵ & <10⁻⁵ & <10⁻⁵ & <10⁻⁵ \\
0.25 & 0.001\% & 0.006\% & 0.001\% & 0.002\% \\
0.30 & 0.011\% & 0.022\% & 0.014\% & 0.049\% \\
0.35 & 0.080\% & 0.154\% & 0.072\% & 0.229\% \\
0.40 & 0.270\% & 0.529\% & 0.226\% & 0.770\% \\
0.50 & 1.244\% & 2.390\% & 1.163\% & 3.560\% \\
\bottomrule
\end{tabularx}
\end{table}

Three findings are worth noting.

First, the design tolerates variability up to $\sigma_{\log} \approx 0.25$ with no observable errors — a 67\% headroom over the nominal value. Commercial HfO₂ RRAM typically achieves $\sigma_{\log} \approx 0.10$--0.20, placing the design comfortably inside the safe region.

Second, the "0" (undefined) state is approximately twice as error-prone as ±1 at any given variability. This is a structural property of any ternary encoding: the middle state is vulnerable to noise excursions in both directions, whereas the extreme states are vulnerable only in one. It is not a flaw of the design — it is a property of three-valued readout under Gaussian noise — but it justifies designing the undefined state to be the \textbf{safe default}: an assertion drifting from +1 or −1 toward 0 becomes "unanswerable," which a clinical controller can flag for refresh, rather than silently flipping to the opposite truth value.

Third, cascade errors scale roughly as 2--3× the single-junction error rate over a 4-step cascade, not exponentially. This is the benefit of threshold regeneration (Section~\ref{sec:cascade}): each cycle restores the signal to a clean ternary state before feeding the next. Without regeneration, analog accumulation over 4 steps would reduce the noise margin by roughly $\sigma \times \sqrt{4} = 2\sigma$, leading to much faster degradation. With regeneration, the compounding is approximately linear in cycle count rather than multiplicative.

\textbf{What the simulation does not cover.} The simulation is behavioral, not circuit-level. It does not model transistor dynamics (the 10 ns cycle estimate is a timing budget, not a SPICE result), long-term resistance drift (months to years), write endurance (10⁶–10⁹ cycles for RRAM, not a concern for yearly-update knowledge bases), or full-chapter integration beyond the respiratory subset. These are acknowledged as open items for tape-out validation (§7.4).

\begin{figure}[H]
\centering
\begin{tikzpicture}
\begin{groupplot}[
  group style={group size=2 by 1, horizontal sep=1.8cm},
  width=0.44\textwidth,
  height=0.30\textwidth,
  xmin=0.10, xmax=0.50,
  xtick={0.10,0.15,0.20,0.25,0.30,0.35,0.40,0.50},
  xlabel={$\sigma_{\log}$},
  ymode=log,
  ymin=1e-5, ymax=1e1,
  ylabel={Error rate (\%)},
  grid=both,
  major grid style={draw=gray!35},
  minor grid style={draw=gray!15},
  tick align=outside,
  legend style={draw=none, fill=none, font=\small},
  every axis plot/.append style={thick, mark=*, mark size=1.8pt},
]

\nextgroupplot[
  title={Single-junction readout},
  legend pos=north west,
]
\addplot[color=blue] coordinates {
  (0.10,1e-5) (0.15,1e-5) (0.20,1e-5) (0.25,0.001) (0.30,0.011) (0.35,0.080) (0.40,0.270) (0.50,1.244)
};
\addlegendentry{$+1$}
\addplot[color=orange] coordinates {
  (0.10,1e-5) (0.15,1e-5) (0.20,1e-5) (0.25,0.006) (0.30,0.022) (0.35,0.154) (0.40,0.529) (0.50,2.390)
};
\addlegendentry{$0$}
\addplot[color=teal!70!black] coordinates {
  (0.10,1e-5) (0.15,1e-5) (0.20,1e-5) (0.25,0.001) (0.30,0.014) (0.35,0.072) (0.40,0.226) (0.50,1.163)
};
\addlegendentry{$-1$}
\addplot[color=black, dashed, mark=none] coordinates {(0.15,1e-5) (0.15,1e1)};

\nextgroupplot[
  title={4-step cascade},
]
\addplot[color=red!75!black] coordinates {
  (0.10,1e-5) (0.15,1e-5) (0.20,1e-5) (0.25,0.002) (0.30,0.049) (0.35,0.229) (0.40,0.770) (0.50,3.560)
};
\addplot[color=black, dashed, mark=none] coordinates {(0.15,1e-5) (0.15,1e1)};
\end{groupplot}
\end{tikzpicture}
\caption{Error rate versus device variability at SNR = 20 dB: left, per-state single-junction error; right, 4-step cascade error. The dashed line marks the nominal operating point. Values shown as $<10^{-5}$ in the table above are plotted at $10^{-5}$ for visualization.}\label{fig:error-variability}
\end{figure}

\section{Comparison: CDC Chip vs. Neural Accelerator for Medical Reasoning}

\begin{table}[H]
\centering
\small
\begin{tabularx}{\textwidth}{Y|Y|Y}
\toprule
Aspect & Neural Memristive Accelerator & CDC Reasoning Chip \\
\midrule
\textbf{What one junction means} & A weight (fragment, no meaning alone) & A clinical assertion (complete, self-contained) \\
\textbf{Reading one junction gives you} & One term of a dot product & One clinical answer \\
\textbf{Domain scoping} & Impossible at hardware level & Physical array isolation \\
\textbf{Typed inheritance} & Not applicable & Meta-gated inter-array paths \\
\textbf{Cross-axis linkage} & Not applicable & Cross-reference registers \\
\textbf{Consistency checking} & Requires full inference + external logic & Read-before-write on same array \\
\textbf{Transitive classification} & Requires multi-layer + activation + softmax & Cascade reads within array \\
\textbf{Precision per junction} & 4–8 bits (16–256 levels) & 1.58 bits (3 levels) \\
\textbf{Read circuit} & Precision ADC & Two-threshold comparator \\
\textbf{Three-valued logic (true/false/undefined)} & Not representable & Native \\
\textbf{For ICD-11 respiratory diseases} & $\approx$ 10M junctions (embeddings) & $\approx$ 136K junctions (assertions) \\
\textbf{Auditability} & Post-hoc explanation methods & Each junction's address IS the evidence \\
\bottomrule
\end{tabularx}
\end{table}

\section{Discussion}

\subsection{What Is New Here}

Not ternary computing (Knuth, 1969; Soviet Setun, 1958). Not in-memory computing (Hu et al., 2018). Not memristive logic (Borghetti et al., 2010). Not TCAM (Pagiamtzis \& Sheikholeslami, 2006). Not hyperdimensional in-memory computing (Karunaratne et al., 2020). What is new:

\textbf{A structural homomorphism between a materialized CDC domain algebra and a crossbar topology, with an explicit treatment of bridges.} Prior memristive computing maps numerical values to resistance states — the physical layout carries no algebraic meaning. HDC maps vector components to junctions — meaning is distributed. TCAM stores match patterns in flat arrays — entries can be rearranged arbitrarily. This work maps an entire algebra (Heyting lattice with typed Galois connections, established in Paper 2, together with cross-fiber bridges introduced in Paper 3) to a crossbar topology where the wiring IS the algebra. The specialization order is the chip's inter-array connectivity. The meet is a topological common ancestor. The typing function, unfolded across inheritance edges, drives meta-controlled gates. Bridges across non-comparable fibers are cross-reference registers. The closure property is the physical idempotence of re-running inheritance. None of these structural correspondences exist in neural crossbars, HDC arrays, or TCAMs.

\subsection{The Deeper Observation}

Every computing architecture since von Neumann assumes that storage and computation require different physical mechanisms. Memory cells store; logic gates compute; buses connect them. Even "in-memory computing" only partially addresses this: memristive crossbars compute matrix products in memory, but interpretation, control flow, and decision logic remain external.

CDC's domain algebra suggests why this separation persists: \textbf{prior data formats do not natively encode computational semantics.} A 16-bit floating-point weight of 0.7324 carries no inherent semantic meaning — three identical weights in three different networks compute three different things. A ternary state of +1 at position (CA40.00, is\_a, Pneumonia) in the @Anatomical array carries the same meaning no matter what chip it is printed on. The position, the domain, the relation, and the ternary value together fix the semantics; no interpretive layer is needed.

When the data format is natively computational, the storage-computation separation dissolves — not just in theory (Paper 3), but in physics (this paper).

\subsection{A Broader Structural Reading}

The homomorphism Φ admits a broader reading. The C-R@D-C′ form is not unique to CDC: it appears in branching control flow (state → transition@condition → state), in abstract syntax trees, in state-transition machines, and in Kripke accessibility structures. Whether these constitute parallel instantiations of a common structural form, or merely share surface notation, is beyond the scope of this paper. We note only that the storage-computation dissolution demonstrated here in hardware is consistent with the software-layer result of Paper 3, suggesting the dissolution may not be substrate-specific. A characterization of what substrate properties suffice to support the C-R@D-C′ form — and the conditions under which storage-computation separation reemerges — is left for future work.

\subsection{Limitations}

\textbf{Domain-specific.} This chip performs domain-constrained classification and reasoning. It does not do general-purpose computation, neural inference, or numerical simulation.

\textbf{Compile-time materialization.} The domain structure is fixed at fabrication. Runtime evolution of the domain lattice itself is not supported; software-level mechanisms that split or merge domains must generate a new materialization, which maps to a new chip (or a reconfigurable platform — not pursued here). For stable reference standards such as ICD-11 that update on yearly cycles, this boundary is acceptable; for fast-changing knowledge, it is a real limitation.

\textbf{Static knowledge within each materialization.} Reprogramming junctions requires write pulses. For rapidly changing assertions within a fixed domain structure, write endurance becomes relevant. A practical update protocol: freeze the chip during inference; batch-apply standard updates at fixed intervals; version the meta-array to roll back if inconsistency is detected.

\textbf{State drift.} RRAM and PCM devices exhibit resistance drift over months to years. A +1 junction drifting toward $R_{\mathrm{mid}}$ converts a clinical assertion to "undefined" rather than flipping it to the opposite truth value — as noted in §5.4, this is the safe failure direction for medical applications, since the system flags an unanswerable query rather than returning a silently incorrect one. Mitigation: periodic readout-and-refresh cycles; ECC across redundant fiber copies.

\textbf{Interconnect area overhead.} The claim that "the wiring is the algebra" commits the design to routing an irregular domain lattice directly onto 2D silicon metal layers. For the respiratory-chapter scale (47 arrays, tens of thousands of inter-array wires) this is tractable, but as the number of domains grows toward full ICD-11 ($\approx$ 500 domains) or beyond, irregular inter-array routing will dominate both floorplan area and critical-path delay — the dense 1T1R arrays become a small fraction of total silicon. Future scaled designs will likely need to map inter-array connections onto a programmable \textbf{Network-on-Chip (NoC)} fabric or a shared-bus arbitration layer with routing tables held in SRAM, trading pure topological mapping for spatial efficiency. The homomorphism Φ would then hold up to a fixed routing indirection — a structurally unchanged but physically more realistic implementation.

\textbf{Proof-of-concept scale.} The design covers one ICD-11 chapter. Full deployment requires multi-chip integration or larger monolithic arrays, and — as above — likely a NoC-based inter-array fabric.

\textbf{No SPICE validation.} This paper provides a behavioral simulation (Section~\ref{sec:behavioral-simulation}) but not transistor-level SPICE. The 10 ns cycle estimate is a timing budget, not a verified circuit simulation. Read-margin under process corners, sense-amplifier offset characterization, buffer-tree skew, and parasitic RC extraction on the inter-array wiring all remain to be characterized before tape-out.

\section{Conclusion}

We presented a memristive chip design grounded in a structural homomorphism Φ between a materialized CDC domain algebra (Li, Wang \& Zhao, 2026b) and a crossbar topology. The homomorphism preserves fiber isolation as array isolation, the specialization order as directed wiring, meet and Heyting implication as topological properties, ternary valuation as three resistance states, the typed Galois connection (after unfolding τ over inheritance edges) as meta-controlled gates, cross-fiber bridges as cross-reference registers, and the closure property as physical idempotence.

This homomorphism is what separates the design from ternary memory. A TCAM stores values in a flat structure where physical layout is arbitrary. The CDC crossbar topology IS the domain algebra — the wiring encodes the specialization order, the gates encode the typing function, the registers encode the bridges, and rearranging connections changes reasoning semantics. \textbf{This is not a chip that stores medical knowledge. This is a chip whose physical structure is medical knowledge.}

Five reasoning capabilities — domain scoping, three-valued logic, transitive classification, typed inheritance, and write-time consistency checking — follow as direct consequences of the homomorphism, with cross-axis queries handled by bridge register lookups. Each is physically realized: array isolation prevents cross-domain leakage; three resistance states encode holds/negated/undefined; a row-column cascade circuit within each 1T1R array computes transitive closure; meta-gates control selective inheritance; read-before-write checks consistency; cross-reference registers implement bridges. Behavioral simulation at the nominal operating point ($\sigma_{\log}=0.15$, SNR = 20 dB) observes zero errors across all six capabilities in 100,000 trials each, with a safety margin extending to $\sigma_{\log} \approx 0.25$ before any errors appear — a margin that comfortably accommodates commercial RRAM device variability.

The fabrication requirements are modest: three resistance states per junction (vs. 16–256 for neural accelerators), 1T1R cells (standard), two-threshold comparators (vs. precision ADCs), a small SRAM bank for bridges, and $\approx$ 136,000 memristive junctions for the respiratory chapter.

The broader implication: the storage-computation separation that has defined computing architecture since von Neumann persists because prior data formats do not carry algebraic structure. When the data representation has algebraic structure — as CDC's domain Heyting algebra does — that structure can be materialized into physical topology at compile time, and the separation dissolves. In software (Paper 3) and in hardware (this paper).

One junction. One clinical assertion. One inference. No software.

\section{References}

Borghetti, J., Snider, G. S., Kuekes, P. J., Yang, J. J., Stewart, D. R., \& Williams, R. S. (2010). 'Memristive' switches enable 'stateful' logic operations via material implication. \emph{Nature}, 464, 873–876.

Chen, A., Hutchby, J., Zhirnov, V. V., \& Bourianoff, G. (Eds.). (2015). \emph{Emerging Nanoelectronic Devices}. John Wiley \& Sons.

Hu, M., et al. (2018). Memristor-based analog computation and neural network classification with a dot product engine. \emph{Advanced Materials}, 30(9), 1705914.

Kanerva, P. (2009). Hyperdimensional computing: An introduction to computing in distributed representation with high-dimensional random vectors. \emph{Cognitive Computation}, 1(2), 139–159.

Karunaratne, G., Le Gallo, M., Cherubini, G., Benini, L., Rahimi, A., \& Sebastian, A. (2020). In-memory hyperdimensional computing. \emph{Nature Electronics}, 3(6), 327–337.

Knuth, D. E. (1969). \emph{The Art of Computer Programming, Volume 2: Seminumerical Algorithms.} Addison-Wesley.

Kvatinsky, S., Belousov, D., Liman, S., Satat, G., Wald, N., Friedman, E. G., Kolodny, A., \& Weiser, U. C. (2014). MAGIC — Memristor-Aided Logic. \emph{IEEE Transactions on Circuits and Systems II}, 61(11), 895–899.

Li, C., Hu, M., Li, Y., Jiang, H., Ge, N., Montgomery, E., Zhang, J., Song, W., Dávila, N., Graves, C. E., Li, Z., Strachan, J. P., Lin, P., Wang, Z., Barnell, M., Wu, Q., Williams, R. S., Yang, J. J., \& Xia, Q. (2020). Analog content-addressable memories with memristors. \emph{Nature Communications}, 11, 1638.

Li, C., Wang, Y., \& Zhao, C. (2026a). Domain-constrained knowledge representation: A modal framework. arXiv:2604.01770.

Li, C., Wang, Y., \& Zhao, C. (2026b). Domain-contextualized inference: A computable graph architecture for explicit-domain reasoning. arXiv:2604.04344.

Li, C., \& Wang, Y. (2026c). Reasoning as data: Representation-computation unity and its implementation in a domain-algebraic inference engine. arXiv:2604.10908.

Merolla, P. A., et al. (2014). A million spiking-neuron integrated circuit with a scalable communication network and interface. \emph{Science}, 345(6197), 668–673.

Pagiamtzis, K., \& Sheikholeslami, A. (2006). Content-addressable memory (CAM) circuits and architectures: A tutorial and survey. \emph{IEEE Journal of Solid-State Circuits}, 41(3), 712–727.

Talati, N., Gupta, S., Mane, P., \& Kvatinsky, S. (2016). Logic design within memristive memories using Memristor-Aided loGIC (MAGIC). \emph{IEEE Transactions on Nanotechnology}, 15(4), 635–650.

Wong, H.-S. P., \& Salahuddin, S. (2015). Memory leads the way to better computing. \emph{Nature Nanotechnology}, 10(3), 191–194.

Wu, T. F., Li, H., Huang, P.-C., Rahimi, A., Rabaey, J. M., Wong, H.-S. P., Shulaker, M. M., \& Mitra, S. (2018). Brain-inspired computing exploiting carbon nanotube FETs and resistive RAM: Hyperdimensional computing case study. \emph{2018 IEEE International Solid-State Circuits Conference (ISSCC)}, pp. 492–494.

Yao, P., Wu, H., Gao, B., Tang, J., Zhang, Q., Zhang, W., Yang, J. J., \& Qian, H. (2020). Fully hardware-implemented memristor convolutional neural network. \emph{Nature}, 577(7792), 641–646.

Yu, S., Gao, B., Fang, Z., Yu, H., Kang, J., \& Wong, H.-S. P. (2013). A low energy oxide-based electronic synaptic device for neuromorphic visual systems with tolerance to device variation. \emph{Advanced Materials}, 25(12), 1774–1779.

\emph{The algebra is the topology. The topology is the chip. The chip is the reasoning.}

\end{document}